\documentclass[twocolumn,showpacs,preprintnumbers,amsmath,amssymb]{revtex4}
\usepackage{times}
\usepackage{color}
\usepackage{amsfonts}
\usepackage{ulem}
\usepackage{bm}
\newcommand {\LSVO}{La$_{1-x}$Sr$_x$VO$_{3}$}
\def\nn{\nonumber}
\def\beq{\begin{equation}}
\def\eeq{\end{equation}}
\def\kb{k_{\rm B}}
\def\na{n_{\rm A}}
\def\nb{n_{\rm B}}
\def\ga{g_{\rm A}}
\def\gb{g_{\rm B}}

\usepackage[abs]{overpic}

\begin{document}

\title{Temperature Dependence of Thermopower in Strongly Correlated Multiorbital Systems}

\author{M. Sekino$^{1,4}$}
\author{S. Okamoto$^{2}$}
\author{W. Koshibae$^{3,4}$}%
\author{M. Mori$^{1,4}$}%
\author{S. Maekawa$^{1,4}$}

\affiliation{
$^1$Advanced Science Research Center, Japan Atomic Energy Agency, Tokai, Ibaraki 319-1195, Japan\\
$^2$Materials Science and Technology Division, Oak Ridge National Laboratory, Oak Ridge, Tennessee 37831, USA\\ 
$^3$RIKEN Center for Emergent Matter Science (CEMS), 2-1 Hirosawa, Wako,
Saitama 351-0198, JAPAN\\
$^4$JST, CREST, Sanbancho, Tokyo 102-0075, Japan 
}
\date{\today}

\begin{abstract}
Temperature dependence of thermopower in the multiorbital Hubbard model is studied 
by using the dynamical mean-field theory with the non-crossing approximation  
impurity solver. 
It is found that the Coulomb interaction, the Hund coupling, and the crystal filed splitting 
bring about 
non-monotonic temperature dependence of the thermopower, including its sign reversal. 
The implication of our theoretical results to some materials is discussed.
\end{abstract}

\pacs{72.15.Jf, 71.10.Fd, 75.20.Hr}

\maketitle

The thermopower in strongly correlated electron systems has been attracting much attention~\cite{chaikin,terasaki,kosh00,kosh01,maekawabook,zlaticbook,uchida,mari,pruschke1,pruschke2,palsson,merino,oudovenko}. 
The thermopower is 
the entropy flow by the electric current, so that 
the spin and orbital degrees of freedom are 
of importance to enhance the thermopower~\cite{chaikin,kosh00,kosh01,maekawabook}.
In the strongly correlated systems, the thermopower  
often shows the non-monotonic temperature dependence~\cite{uchida,mari,oudovenko}.
Furthermore, in the multiorbital correlated system, 
the electronic state reflects the crystal field splitting 
and the Hund coupling. 
So far, the thermopower in the multiorbital correlated electron system has been studied by several 
groups~\cite{oudovenko,kosh01},  
while 
the effect of the crystal field splitting and the Hund coupling has not been fully examined on the temperature dependence of the thermopower in multiorbital correlated systems.   

In this paper, we study the thermopower in the multiorbital Hubbard model given by,
\begin{eqnarray}
H &=&\sum_{k,m,\sigma} \left(\varepsilon_k + \Delta_m\right) c_{km\sigma}^\dag c_{km \sigma}
+ U\sum_{i,m} n_{im\uparrow} n_{im\downarrow} \nn \\ 
&+& U'\sum_{i,m<m',\sigma,\sigma'} n_{im\sigma} n_{im'\sigma'} \nn \\
&-& J \sum_{i,m<m',\sigma,\sigma'} \left( c_{im\sigma}^{\dagger} c_{im\sigma'} c_{im'\sigma'}^{\dagger} c_{im'\sigma}
      +\mbox{h.c.} \right)  \nn\\
&+& I \sum_{i,m<m'}  \left(c^{\dagger}_{im\uparrow} c^{\dagger}_{im\downarrow} c_{im'\downarrow} c_{im'\uparrow}
      +\mbox{h.c.} \right),
\label{Hamiltonian}
\end{eqnarray}
with the electron spin $\sigma(=\uparrow, \downarrow)$, the orbital index $m$, and 
the dispersion relation of the non-interacting electrons $\varepsilon_k$.  
Here, $\Delta_m$ denotes the energy level of the $m$-th orbital, 
$U$ $(U')$ is the intraorbital (interorbital) Coulomb repulsion, 
$J$ and $I$ are the magnitude of Hund coupling and pair hopping, respectively. 
Below, we impose the conditions, $U=U'+J+I$ and $J=I$ 
assuming the orbital symmetry, $t_{2g}$ or $e_g$.

The many-body interactions are dealt with using 
the dynamical mean-field theory (DMFT)~\cite{uchida,mari,georges,pruschke2,pruschke1,palsson,merino,oudovenko}
 with the non-crossing approximation (NCA)~\cite{maekawa,bickers,pruschke1}  
impurity solver.  
In this work, we consider the semicircular density of states 
for the non-interacting system, 
\beq
D_0(\omega)=\frac{2}{\pi W_0^2} 
	\sqrt{W_0^2 - \omega^2}.
\eeq 
The thermopower $Q$ is then given by
$
Q=-(\kb/e) (A_1/A_0) 
$ 
with, 
\begin{eqnarray}
A_l=\frac{\pi}{\hbar \kb} \sum_{m,\sigma} \int d \omega d \epsilon 
	\frac{ (\beta \omega)^l}{4\cosh^2 \left(\beta \omega /2 \right)} 
	\left[\rho_{m,\sigma}(\omega, \epsilon)\right]^2 D_0(\epsilon),~~  
	\label{AlD0}
\end{eqnarray}  
where $\beta=1/\kb T$~\cite{pruschke2,palsson} . 
The spectral density $\rho_{m,\sigma}(\omega, \epsilon)$  
is given by
$\rho_{m,\sigma}(\omega, \epsilon)\equiv -\mbox{Im} G_{m,\sigma}(\omega, \epsilon)/\pi$ 
with the Green's function $G_{m,\sigma}(\omega, \epsilon)$ obtained by DMFT. 
Below, $W_0$ is taken to be unity.  

\begin{figure}
\begin{center}
\begin{overpic}[width=8cm]{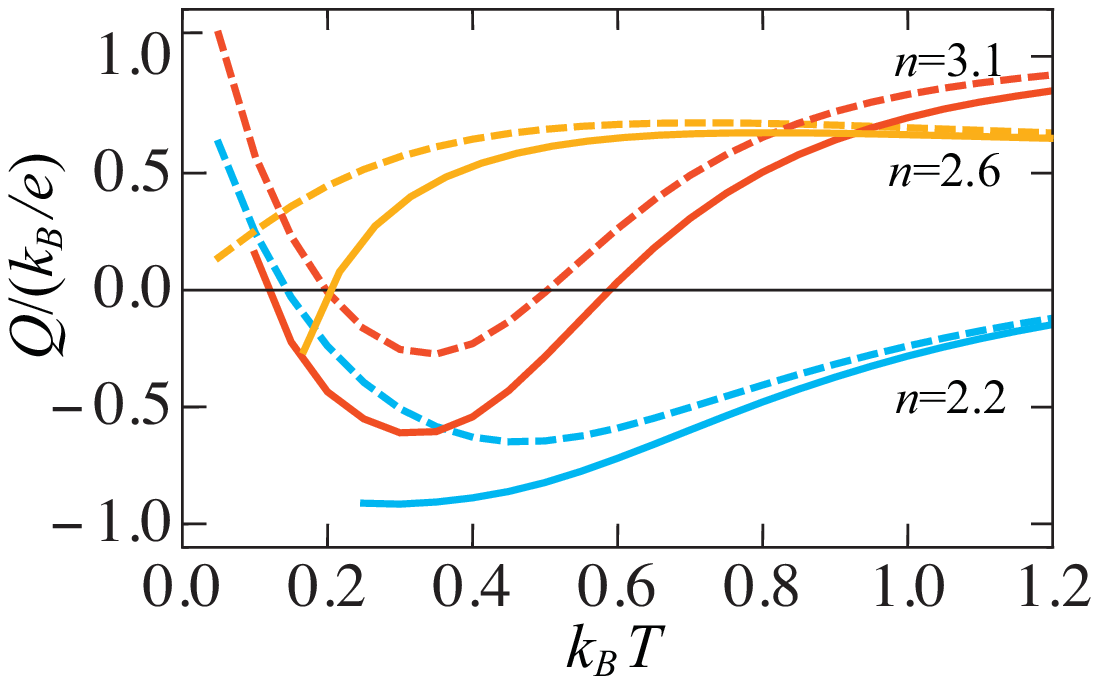}
\put(-10,130){(a)}
\end{overpic}
\end{center}
\begin{center}
\begin{overpic}[width=8cm]{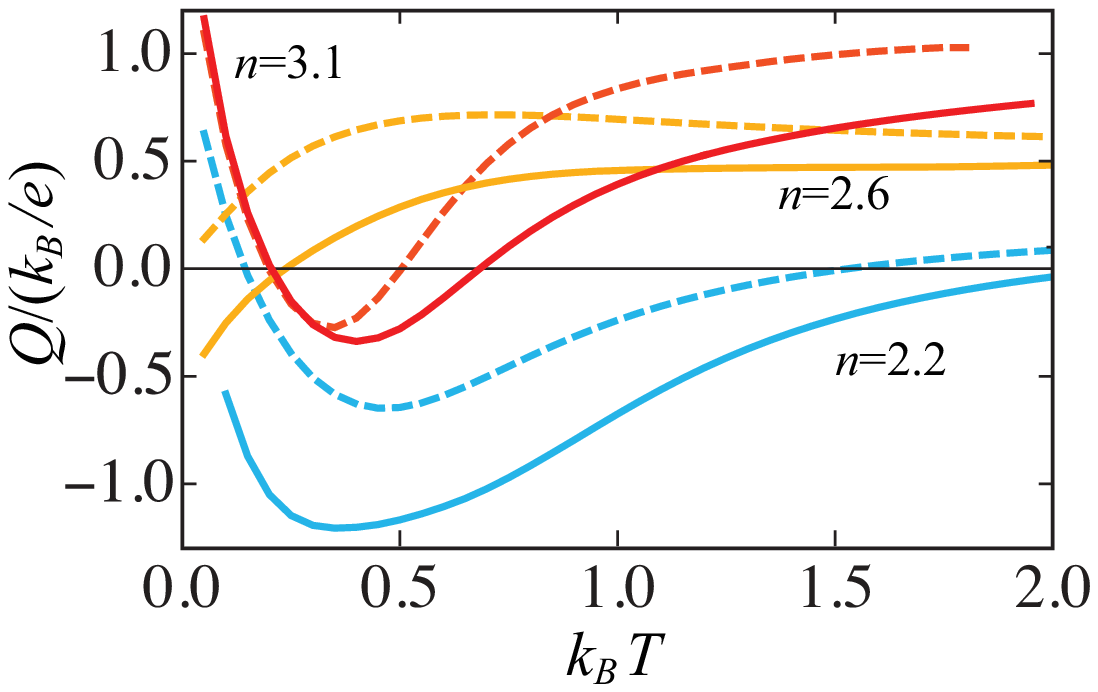}
\put(-10,130){(b)}
\end{overpic}
\end{center}
\caption{(Color online) 
(a) Temperature dependence of the thermopower for $U=3$ and $J=0$ with $n$=2.2, 2.6, and 3.1. 
The broken (solid) line indicates the thermopower for $\Delta=0$ ($\Delta=0.4$).  
(b) Temperature dependence of the thermopower for $U=3$, and $\Delta=0$ with $n$=2.2, 2.6, and 3.1. 
The broken (solid) line represents the thermopower for $J=0$ ($J=0.5$).} 
\label{fig1}
\end{figure}

First, we examine the two-orbital system with $\Delta_1=-\Delta_2=-\Delta/2$.  
Figures \ref{fig1} (a) and (b) show the effect of crystal-field splitting $\Delta$ and $J$ 
on the temperature dependence of 
$Q$, 
respectively~\cite{foot}.  
As in the single-orbital model~\cite{mari}, we find the non-monotonic temperature dependence of the thermopower due to the Coulomb interaction. 
This behavior resembles the experimental report on doped vanadates~\cite{uchida}. 
By introducing finite $\Delta$ and/or $J$, 
all the lines are shifted down (see the broken and solid lines in Fig.~\ref{fig1}). 

Here, we briefly summarize the Heikes formula~\cite{kosh00,kosh01,maekawabook,chaikin} 
being useful to understand the non-monotonic temperature dependence and the shifts by $\Delta$ and/or $J$. 
The following two cases are considered in the high-temperature limit; 
(1) $Q_1= Q(T\rightarrow\infty,U)$ by keeping $\kb T< U$, and 
(2) $Q_2= Q(T\rightarrow\infty,U)$ by keeping $\kb T> U$.
In the case (1), $U\rightarrow\infty$ is taken before $T\rightarrow\infty$.  
In this limit, except for integer fillings, 
one needs to consider two kinds of sites, $A$ and $B$, 
 where the electron occupancy in $A$-site, $\na$, differs from that in $B$-site, $\nb$, by one, i.e., $\na-\nb=1$. 
The each sites have the local degeneracies, $\ga$ and $\gb$, in accordance with the electron occupancy.  
Therefore, the high-temperature limit $Q_1$ is given by,
\beq
Q_1=-\frac{\kb}{e}\ln\frac{\ga}{\gb}-\frac{\kb}{e}\ln\left(\frac{n-\na}{\nb-n}\right),\label{q1}
\eeq 
where $n$ is the electron density with $\na>n>\nb$.   
In the case (2), on the other hand, $U$ is of less importance, so that $Q_2$ in the $\nu$-orbital system is given by,
\beq
Q_2=-\frac{\kb}{e}\ln\left(\frac{2\nu -n}{n}\right). \label{q2}
\eeq 
Note that 2 in Eq. (\ref{q2}) is associated with the spin degree of freedom. 

The non-monotonic temperature dependences shown in Fig.~\ref{fig1} 
are understood by the change 
from $Q_1$ to $Q_2$.   
As an example, let us consider the temperature dependence of $Q$ for $n=3.1$ with $U=3$, $J=I=0$ and $\Delta=0$ in Fig.~\ref{fig1}(a) (see the red broken line).  
In this case, 
$(\na, \nb)=(4,3)$ and $\ga/\gb=1/4$ result in negative $Q_1$, 
while $Q_2$ is positive.  
For the temperature region $\kb T<U$, the Coulomb interaction is in effect  
for the entropy transport, and hence, 
$Q$ approaches $Q_1$~\cite{chaikin,kosh00,kosh01,maekawabook,uchida,mari}.    
On the other hand, 
at high-temperatures $\kb T \gg U$, , 
the effect of $U$ is of less importance, so that 
$Q$ approaches $Q_2$.  
The change from $Q_1$ to $Q_2$ is seen 
in the temperature dependence of $Q$ in Fig.~\ref{fig1}(a), i.e., 
the broken line for $n=3.1$ changes the sign from negative to positive 
at the temperature region, $0.3<\kb T$.  
This picture consistently explains the other cases shown by broken lines in Fig.~\ref{fig1}.  

By including $\Delta$, $Q$ is shifted down in the negative direction as shown in Fig.~\ref{fig1}(a) 
(see the difference between the broken and solid lines).  
For finite $\Delta$, electrons prefer to go into the lower-energy-level orbital. 
Let us consider the effect of $\Delta$ in the formula $Q_1$.  
It is noted that the first term of $Q_1$, i.e., $-(\kb/e)\ln(\ga/\gb)$ 
decreases for $\Delta\rightarrow\infty$ in all the cases in Fig.~\ref{fig1}(a).  
(For example, for $2<n<3$, $\ga/\gb=4/6$ for $\Delta=0$, 
while $\ga/\gb=2/1$ 
for $\Delta\rightarrow\infty$.)  
Hence, $Q_1$ is shifted down in the negative direction.  

Next, we discuss the effect of Hund coupling on the thermopower based on $Q_1$ (see Fig.~\ref{fig1}(b)). 
Note that the condition $U'-J=U-3J$ (=3) is imposed to roughly fix the Mott gap.  
Consequently, the magnitude of the on-site Coulomb interaction $U$ increases 
by including a finite $J$.  
Therefore, the temperature region 
where the Coulomb interaction is in effect on $Q$ is expanded, 
and hence the lines in Fig.~\ref{fig1}(b) 
are shifted down in the high temperature region~\cite{mari}.  
In addition to the shift by $U$, 
the lines for $n=2.2$ and $n=2.6$ in Fig.~\ref{fig1}(b) 
are shifted down particularly around $\kb T\sim 0.5$.  
The shifts are explained by the 
change of spin degeneracy by 
$J$.  
For $n=2.2$ and $n=2.6$, i.e., $2<n<3$ with $\Delta$=0, 
$\ga/\gb=4/6$ for $J$=0, while 
$\ga/\gb=4/3$ for $J\rightarrow \infty$.
This leads to the change in $Q_1$ and then
the shifts of the lines for $n=2.2$ and $n=2.6$ 
in Fig.~\ref{fig1}(b) around $\kb T\sim 0.5$. 
For $3<n<4$, on the other hand, 
the degeneracies $\ga$ and $\gb$ are not modified by $J$.   
Consequently, the result for $n=3.1$ 
around $\kb T\sim 0.5$ 
is not much affected by $J$ in comparison with 
the other electron densities at $2<n<3$. 
The consideration on $Q_1$ and $Q_2$ consistently 
explains the effects of $J$ on the calculated results shown in Fig.~\ref{fig1}(b).

As a summary of the two-orbital model, 
the discussion on $Q_1$ and $Q_2$ 
gives the following understanding 
for the temperature dependence of thermopower:    
(i) The strong Coulomb interaction gives 
the non-monotonic temperature dependence, i.e., the change from $Q_1$ to $Q_2$.  
(ii) The effect of crystal field splitting and Hund coupling is explained by 
the first term of $Q_1$.  

\begin{figure}[t]
\begin{center}
\includegraphics[width=8cm]{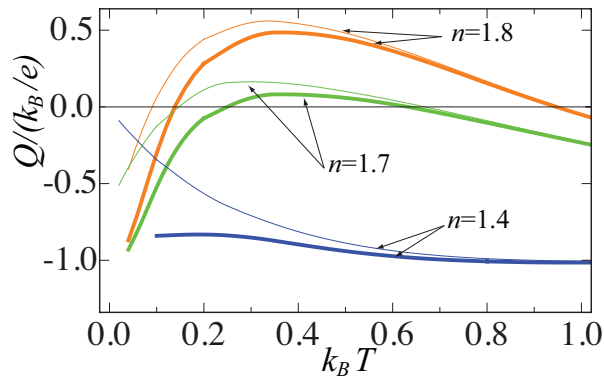}
\end{center}
\caption{(Color online)
Temperature dependence of the thermopower in the three-orbital Hubbard model for
$U'-J=3$ and $J=0.5$ with $n$=1.4, 1.7, and 1.8. 
The thin and thick lines are for $\Delta=0$ and $\Delta=0.25$, respectively. 
}
\label{fig2}
\end{figure}

Finally, 
we consider 
the three-orbital model 
with $\Delta_1=\Delta_2=-\Delta_3=\Delta$, and the electron densities $1<n<2$. 
Figure~\ref{fig2} shows the temperature dependence of $Q$ in 
the three-orbital model for $n$=1.4, 1.7, and 1.8~\cite{foot}. 
Here, we take the parameter set, $U'-J=3$ and $J=0.5$.  
The consideration on $Q_1$ and $Q_2$ is again useful to understand 
the non-monotonic temperature dependence and 
the effect of $\Delta$.  
For example, for $n=1.8$ with $\Delta=0$, 
$(\na,\nb)=(2,1)$ and $\ga/\gb=9/6$, and therefore $Q_1$ is positive, while $Q_2$ is negative. 
This explains the temperature dependence of $Q$ for this doping. 
By introducing $\Delta$, the first term of $Q_1$, $-(\kb/e)\ln(\ga/\gb)$, 
``decreases'' in all the cases in Fig.~\ref{fig2} 
because $\ga/\gb$ is increased to $6/2$ as $\Delta\rightarrow\infty$.  
Hence, all the lines shown in Fig.~\ref{fig2} 
are shifted down in the negative direction. 
 
In our previous study~\cite{uchida}, 
we have reported the non-monotonic temperature dependence 
of the thermopower in \LSVO, which is well explained by 
considering $Q_1$ and $Q_2$ in the three orbital system. 
In this material, the vanadium valence is in between +3 ($d^2$) and +4 ($d^1$) 
and the transport properties are determined by the $t_{2g}$ manifold associated with the three orbital system. 
In this study, 
we have clarified the effect of $\Delta$ and $J$  
on the temperature dependence of the thermopower.  
In reality, the crystal field splitting can be modified by applying a pressure. Thus, high pressure measurements are highly desired to test our theoretical results.   

We have shown the temperature dependences of the thermopower 
in the two- and three-orbital Hubbard models 
by using the dynamical mean-field theory with 
the non-crossing approximation
impurity solver. 
It has been clarified how the Hund coupling, 
the crystal field splitting and the Coulomb interaction 
produce the non-monotonic temperature behavior 
of the thermopower.  
It is also found that the sign of thermopower 
is changed by temperature and electron density. 
The entropy consideration at high temperatures, 
i.e., Heikes formula, consistently explains 
the effect of the crystal field splitting and the Hund coupling 
on the thermopower.    

We would like to thank to G. Khaliullin and V. Zlati\'{c} for useful discussions.
This work is partly supported by Grants-in-Aid for Scientific Research from MEXT 
(Grant No. 24540387, No. 24360036, No.23340093, and No.22340097), JST-CREST, and FIRST-Program. 
S.O. was supported by the U.S. Department of Energy, Basic Energy Sciences, Materials Sciences and Engineering Division.


\begin{thebibliography}{9}
\bibitem{foot}
The numerical calculation of the thermopower 
on DMFT with NCA impurity-solver becomes 
less accurate as $\kb T\rightarrow 0$, i.e., 
the numerical results at 
the temperature region below the local minimum or maximum of the 
temperature dependence of the thermopower are not plausible.
\end{thebibliography}
\end{document}